# Nanomorphology of Annealed P3HS and P3HS:PCBM Films for OPV Applications

Samuele Lilliu[a,b]*, Mejd Alsari[a], Marcus S. Dahlem[a], J. Emyr Macdonald[b]

[a]*Masdar Institute of Science and Technology, Abu Dhabi, UAE*
[b]*School of Physics and Astronomy, Cardiff University, Queens Buildings, The Parade, Cardiff CF243AA, UK*

**Abstract**

Atomic Force Microscopy (AFM) and Grazing Incidence X-Ray Diffraction (GI-XRD) are used to characterize the nanomorphology of spin-coated low (LMW, Mn = 12 kg/mol, regioregularity RR = 84%) and high (HMW, Mn = 39 kg/mol, RR = 98%) molecular weight poly(3-hexylselenophene) (P3HS) films and blend films of P3HS with [6,6]-phenyl-C61-butyric acid methyl ester (PCBM), before and after thermal annealing at 250 °C.

*Keywords:* Organic Solar Cells, Atomic Force Microscopy, Synchrotron X-Ray Diffraction, GI-WAXS

## 1. Introduction

Our previous in-situ synchrotron Grazing-Incidence X-Ray Diffraction (GI-XRD) studies have shown the connection between the increase in poly(3-hexylthiophene) (P3HT) crystallization during annealing and the improvement in Organic Photovoltaic (OPV) devices Power Conversion Efficiency (PCE) [1]. A general path for an extra improvement of the PCE relies on the engineering of polymers with energy bandgap $E_{gap}$ lower than $E_{gap,P3HT}$ [2, 3]. P3HS resembles P3HT, but the S atom is replaced with a Se atom. This gives a lower $E_{gap}$ while keeping the $V_{oc}$ unchanged [2]. However, the Fill Factor (FF) and PCE are lower in P3HS:PCBM OPVs. The study of the nanomorphology of P3HS films conducted here could help to understand the lower performance observed in P3HS based devices.

## 2. Experimental

P3HS was synthesized at the Imperial College of London and subsequently purified to obtain a low molecular weight fraction (Mn = 12 Kg/mol, regioregularity 84%), and a higher weight fraction (Mn = 39 Kg/mol, regioregularity 98%) [4]. For each P3HS batch, pure P3HS and P3HS: PCBM blends (1:1) were prepared inside a nitrogen filled glovebox, by diluting the polymer and fullerene powders in chlorobenzene (30mg/mL) and stirring them overnight (at 50°C). The solutions were then spin-coated onto Si/ SiO$_2$ substrates, giving an approximate film thickness of 50nm. Grazing Incidence X-Ray Diffraction (GI-XRD) measurements were performed at the XMaS beamline (ESRF, Grenoble), with a beam energy of 10 keV and a MAR SX-165 2D detector (images size $q_{xy}$=2 Å$^{-1}$, $q_z$=2 Å$^{-1}$ [5]), with the Cardiff annealing chamber [5-8]. GI-XRD analysis was conducted with Dr. Lilliu's Matlab software [5-8]. Atomic Force Microscopy (AFM) scans were performed with a Nanoscope III (Bruker).

* Corresponding author. Tel.: +97 1 052 947 0884
  *E-mail address:* samuele_lilliu@hotmail.it, slilliu@masdar.ac.ae

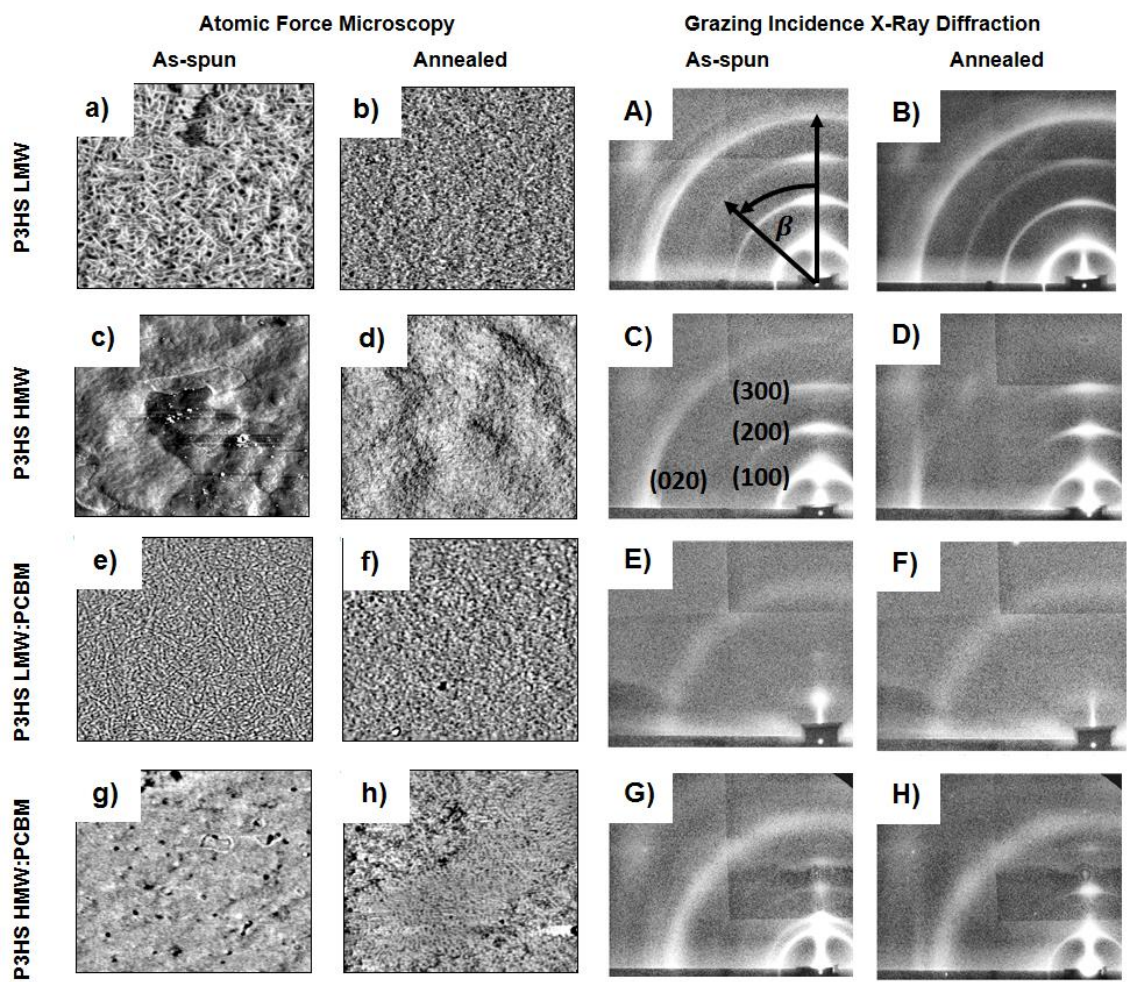

Figure 1 - As-spun and annealed (at 250 °C) P3HS and P3HS:PCBM films with different molecular weights. The AFM scans (a-h) are performed at room temperature with a Nanoscope III (Bruker), in tapping mode with super-sharp tips (scan size 1×1 μm$^2$). The synchrotron GI-XRD scans (A-H) are performed with a beam energy of 10 keV. $β$ indicates the polar angle[5].

## 3. Results and Discussion

Figure 1(a-h) shows a comparison between AFM phase images of as-spun and annealed (*ex-situ* at 250 °C) P3HS and P3HS:PCBM films measured at room temperature. Figure 1(A-H) shows a comparison between GI-XRD scans of as-spun and annealed (*in-situ* at 250 °C) of the same films under the same conditions[4]. In general, P3HS crystallite domain sizes dramatically increase upon annealing to the P3HS melting temperature [4]. The as-spun LMW P3HS exhibits whiskers (or ribbons) of ~100 nm length and ~4 nm height (Figure 1(a)). Annealing gradually disrupts the whiskers, which completely dissolve at ~100 °C (Figure 1(b)). For the HMW P3HS, the as-spun film shows uniform plateaus (Figure 1(c)) that disappear upon annealing at temperatures above 140 °C (Figure 1(d)). Similar differences between LMW and HMW P3HT film morphologies have previously been reported [9, 10]. Simple

models assume that P3HT whiskers are made of edge-on polymer chains stacked along the **b**-direction [11]. However, in our diffraction patterns, we observe a modest polar angular (*β*) dependence in the (100) ring (Figure 1(A-B)) of the LMW P3HS [4].

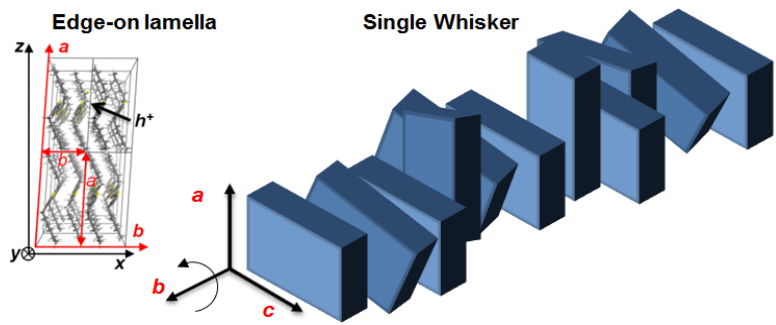

Figure 2 - Illustration showing the possible orientation of a single whisker. The P3HS unit cells are enclosed by the boxes and are described by the lattice vectors *a, b, c*. The P3HT alkyl- and π-stacking directions are along the *a* and *b* vectors, respectively.

It is therefore possible that ribbons still stack along the π-direction, but with a twist about the **b**-direction, as shown by Liu et al. [11] and illustrated in Figure 2. In the case of HMW P3HS, as suggested by GI-XRD data, the plateaus are made of sheets of P3HS edge-on crystalline domains relatively well connected. Features that could be identified as whiskers can also be observed in the LMW P3HS blend (Figure 1(e-f)), but with a lower level of details than in the pure P3HS (Figure 1(a-b)). The GI-XRD data suggest that PCBM could disrupt the polymer crystalline packing in the LMW P3HS blend (Figure 1(E-F)) [4]. These features disappear upon annealing. The as-spun HMW P3HS blend (Figure 1(g-h)) shows several dark spots, likely due to PCBM aggregates. After annealing at temperatures above 200 °C, nano-domains of P3HS in a nano-fibril like shape are clearly visible. This is likely attributed to the domain size growth along the ***a***-direction.

## 4. Conclusion

Finally, our work suggests that differences in $M_n$ and/or RR do not affect the domain size significantly. However, differences in $M_n$ and/or RR affect: (i) the domain orientation and (ii) the impact of the annealing on the amount of crystalline material at different temperatures [4].

### Acknowledgements

We thank Dr. E. Pires, Dr. T.Agostinelli, Prof. J. Nelson, Dr. Mohammed Al-Hashimi, Dr. Martin J. Heeney for providing us with the materials and for the help with the measurements at the beamline.